\documentclass[aps,pre,twocolumn,groupedaddress,showpacs]{revtex4}
\usepackage{graphicx}%
\usepackage{dcolumn}
\usepackage{amsmath}   
\usepackage{bm}
\usepackage{longtable}
\begin{document}

\title{\centering\Large\bf Cavity Field in Molecular Liquids. 
                           When a Polar Liquid Becomes a Dielectric?} 
\author{Daniel R.\ Martin}
\author{Dmitry V.\ Matyushov}
\affiliation{Center for Biological Physics, Arizona State University, 
PO Box 871604, Tempe, AZ 85287-1604}

\date{\today}
\begin{abstract}
  We present the results of an analytical theory and simulations of
  the field inside a cavity created in a dipolar liquid placed in a
  uniform external electric field. The analytical theory shows that
  the limit of continuum electrostatics is reached through a
  singularity in the microscopic response function responsible for a
  non-decaying longitudinal polarization wave. Fields in microscopic
  cavities are much different from macroscopic predictions, and
  low-polarity dielectrics are predicted to have a continuum limit
  distinct from the solution of Maxwell's equations. Computer Monte
  Carlo simulations never reach the standard continuum limit and
  instead converge to the new continuum solution with increasing
  cavity size.
\end{abstract}
\maketitle

Electric fields within cavities in uniformly polarized dielectrics are
commonly calculated by the rules of macroscopic electrostatics
\cite{Landau8}.  In case of an empty spherical cavity carved from a
polarized dielectric, the connection between the field inside the
cavity, $\mathbf{E}_c$, and the macroscopic (Maxwell) field in the
dielectric, $\mathbf{E}$, is particularly simple:
\begin{equation}
  \label{eq:1}
  \mathbf{E}_c = \frac{3\epsilon}{2\epsilon+1}\mathbf{E}=
                 \frac{3}{2\epsilon+1}\mathbf{E}_0 .
\end{equation}
Here, $\epsilon$ is the dielectric constant and $\mathbf{E}_0$ is the uniform
external field. Equation (\ref{eq:1}) is widely used for problems
related to inserting nonpolar \cite{Boettcher:73} and polar
\cite{KumarPRL:03} impurities into dielectrics and, more recently, for
electrical and optical properties of nanopartices in polar matrices
\cite{Kreibig:95}. More generally, the calculation of fields within
cavities in systems with dipolar interactions is fundamental for the
mean-field formulation of susceptibility (magnetic or dielectric) of
such media \cite{Huke:04}. Despite its importance, the limits of the
applicability of Eq.\ (\ref{eq:1}) have never been studied. In
particular, one wonders at which cavity size the laws of macroscopic
electrostatics cease to apply and one needs to deal with microscopic
electric fields. The solution of Maxwell's equations is sensitive to
the macroscopic shape of the dielectric samples \cite{Boettcher:73}
and, given the short length of correlation decay in liquids
\cite{SPH:81}, this picture may break down at a microscopic
length-scale.

Equation (\ref{eq:1}) lends itself directly to tests since it predicts
two physically significant results: (i) the cavity field is
independent of the cavity radius, i.e.\ the cavity length-scale does
not enter the final result, and (ii) the dielectric is infinitely
polarizable, i.e.\ the internal field of polarized dipoles within the
dielectric screens the external field and essentially no field is
expected inside a cavity in dielectrics with high $\epsilon$.  This Letter
analyzes these predictions by using analytical formulation for the
microscopic dielectric response and Monte Carlo (MC) simulations of
cavity fields created inside the model fluid of dipolar hard spheres
(DHS).

A cavity within a dielectric can be described by excluding the
polarization field from its volume \cite{Li:92,Chandler:93}. The
generating functional of the Gaussian polarization field $\mathbf{P}$
can then be written as
\begin{equation}
\label{eq:2}
G(\mathbf{A}) = \int  e^{\mathbf{A}*\mathbf{P} -\beta H_{\text{B}}[\mathbf{P}]}\prod_{\Omega_0} 
\delta\left[\mathbf{P}(\mathbf{r})\right] \mathcal{D}\mathbf{P}.
\end{equation}
The product of $\delta$-functions in Eq.\ (\ref{eq:2}) excludes the
polarization field from the cavity volume $\Omega_0$.  The asterisks
between vectors denote both the space integration and tensor
contraction, and the bath Hamiltonian $H_{\text{B}}[\mathbf{P}]$
describes Gaussian fluctuations of the isotropic polar liquid
characterized by the response function $\bm{\chi}_s(\mathbf{k})$.  In
dipolar liquids with axial symmetry, this response function expands
into longitudinal (L) and transverse (T) projections \cite{Madden:84}
\begin{equation}
  \label{eq:3}
  \bm{\chi}_s(\mathbf{k}) = 
         \frac{3y}{4�} \left[S^L(k) \mathbf{J}^L + S^T(k)\mathbf{J}^T \right],
\end{equation}
where $\mathbf{J}^L=\mathbf{\hat k}\mathbf{\hat k}$ and $\mathbf{J}^T=
\mathbf{1} - \mathbf{\hat k}\mathbf{\hat k}$ are two orthogonal dyads
and $S^{L,T}(k)$ are the structure factors which depend only on the
magnitude of the wave-vector $k$; $y=(4\pi/9)\beta \rho m^2$ is the standard
density of dipoles $m$ usually appearing in dielectric theories, $\rho$
is the number density, and $\beta$ is the inverse temperature.

The constraint imposed on the polarization field to vanish from the
cavity breaks the isotropic symmetry of the system and produces a
non-local response function $\bm{\chi}(\mathbf{k}_1,\mathbf{k}_2)$
(2-rank tensor) depending on two wavevectors \cite{DMjcp1:04}. This
function is obtained by taking the second derivative of
$\ln[G(\mathbf{A})]$ in the auxiliary field $\mathbf{A}$ in Eq.\
(\ref{eq:2}) and setting $\mathbf{A}=0$. The result is
\cite{DMjcp1:04}:
\begin{equation}
  \label{eq:4}
  \bm{\chi}(\mathbf{k}_1,\mathbf{k}_2) = \bm{\chi}_s(\mathbf{k}_1)\delta_{\mathbf{k}_1,\mathbf{k}_2} - 
  \bm{\chi}^{\text{corr}}(\mathbf{k}_1,\mathbf{k}_2),
\end{equation}
where the correction term
$\bm{\chi}^{\text{corr}}(\mathbf{k}_1,\mathbf{k}_2)$ accounts for the
effect of the cavity excluding the polarization field from its volume.

\begin{figure}
  \centering
  \includegraphics*[width=5.8cm]{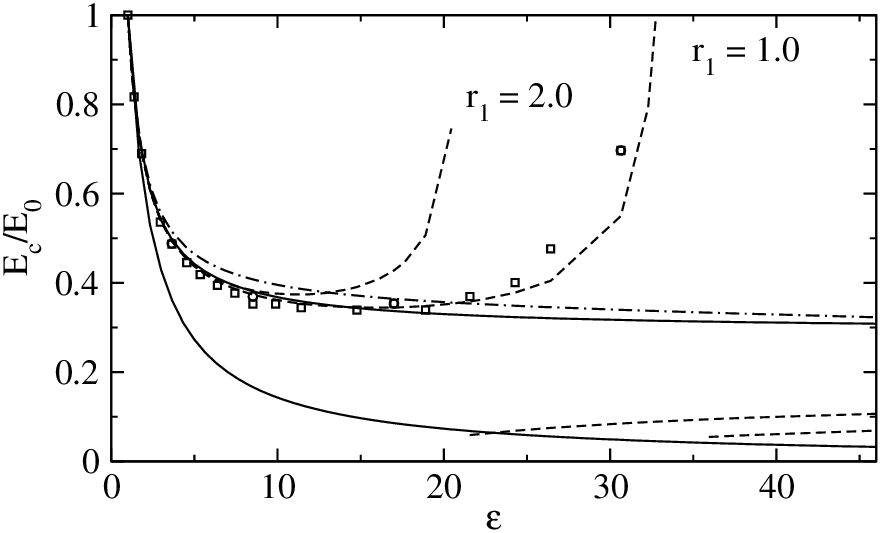}
  \caption{Cavity field calculated from Eq.\ (\ref{eq:7}) with two
    cavity sizes indicated by the distance of closest approach
    $r_1=R_0/ \sigma +0.5$ in the plot. The points where obtained by
    numerical integration in Eq.\ (\ref{eq:7}) with $S^{L,T}(k)$ from
    MC simulations ($r_1=1.0$), while the dashed lines refer to the
    use of analytical $S^{L,T}(k)$ from Ref.\ \onlinecite{DMcp:06}.
    The integral is calculated numerically before the appearance of
    the singularity on the real axis [Eq.\ (\ref{eq:11})] and by
    summation over the poles when the singularity falls on the axis.
    The two methods give identical results when numerical integration
    is justified. The upper and lower solid lines refer to two
    continuum limits, Eq.\ (\ref{eq:10}) and Eq.\ (\ref{eq:1}),
    respectively. The dash-dotted line refers to the lattice summation
    [Eq.\ (\ref{eq:11-1})] instead of continuous integration in Eq.\
    (\ref{eq:7}) taken for a cubic cell of $N=108$, $r_1=2.0$.  }
  \label{fig:1}
\end{figure}

We now consider a spherical cavity of radius $R_0$ inside a
microscopic dielectric liquid and use the response function from Eqs.\
(\ref{eq:3}) and (\ref{eq:4}) to determine the cavity field. For a
dielectric in the uniform external field $\mathbf{E}_0$, the
projection of the field inside the cavity on the direction
$\mathbf{\hat e}_0=\mathbf{E}_0/ E_0$ becomes:
\begin{equation}
  \label{eq:5}
  E_c = E_0 + \mathbf{\hat e}_0\cdot \mathbf{\tilde T}*\bm{\chi}*\mathbf{\tilde E}_0 \cdot \mathbf{\hat e}_0 . 
\end{equation}
Here, $\mathbf{\tilde E}_0 = \mathbf{E}_0\delta_{k,0}$ is the Fourier
transform of the external field and $\mathbf{\tilde T}$ is the
$\mathbf{k}$-space dipole-dipole interaction tensor excluding the hard
cavity core with the radius of closest approach $R_1=R_0+\sigma/2$ ($\sigma$ is
the hard-sphere diameter):
\begin{equation}
  \label{eq:6}
  \mathbf{\tilde T} = -4\pi \mathbf{D}_{\mathbf{k}}\frac{j_1(kR_1)}{kR_1} .
\end{equation}
In Eq.\ (\ref{eq:6}), $\mathbf{D}_{\mathbf{k}}=3\mathbf{\hat k}\mathbf{\hat
  k}-\mathbf{1}$, $\mathbf{\hat k}=\mathbf{k}/k$, and $j_n(x)$ is the
spherical Bessel function of order $n$. After some algebra, one
arrives at the following equation
\begin{equation}
  \label{eq:7}
   \begin{split}
  &\frac{E_c}{E_0} = \frac{\epsilon+2}{3\epsilon} - \frac{4R_1}{3\pi} \frac{\epsilon-1}{\epsilon} \\
    &\times \int_0^{\infty} j_1^2(kR_1)\left(\frac{S^T(k)}{S^T(k)-A(k)} - \frac{S^L(k)}{S^L(k) +2A(k)}\right) dk ,
  \end{split}
\end{equation}
where
\begin{equation}
  \label{eq:8}
  A(k) = \frac{(\epsilon-1)^2}{3\epsilon y} \frac{j_1(2kR_1)}{2kR_1} . 
\end{equation}

Equation (\ref{eq:7}) is the central result of our analytical model.
The first term in Eq.\ (\ref{eq:7}) is the local Lorentz field
\cite{Boettcher:73} which appears in our formalism as the field inside
small cavities \cite{Duan:05} of the size much smaller than the length
of dipolar correlations in the liquid. The opposite limit of
macroscopically large cavities turns out to be harder to derive.

\begin{figure}
  \centering
  \includegraphics*[width=5.8cm]{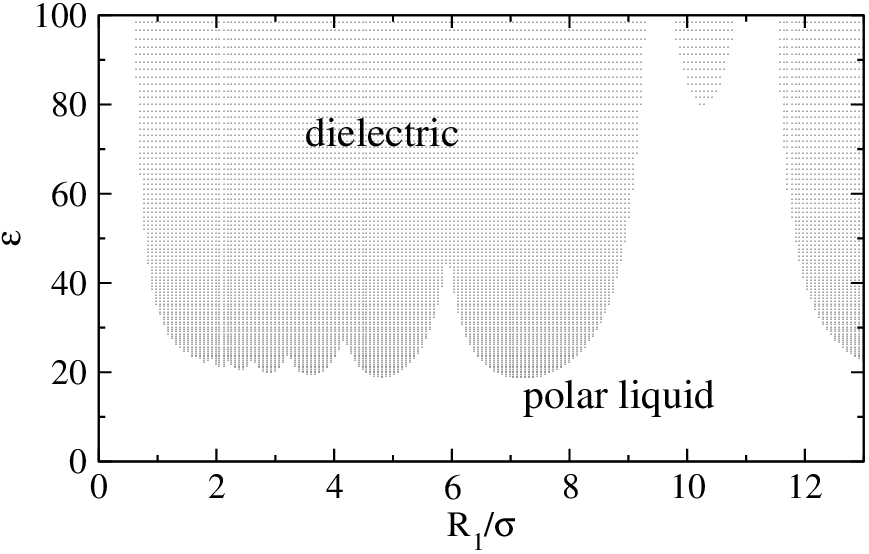}
  \caption{Shaded area indicates the region of existence of a real-$k$
    singularity $k^*$ in the plane of the distance of closest approach
    $R_1 = R_0 + \sigma/2$ and the solvent dielectric constant $\epsilon$. The
    analytical form of $S^L(k)$ (Ref.\ \onlinecite{DMcp:06}) was
    used to solve Eq.\ (\ref{eq:11}) at $\rho^*=0.8$. }
  \label{fig:2}
\end{figure}

The macroscopic (continuum) limit corresponds to the neglect of the
$k$-dependence in the correlation functions representing dipolar
fluctuations of the polar solvent.  If all the functions $S^{L,T}(k)$
and $A(k)$ in the parenthesis under the $k$-integral in Eq.\ (\ref{eq:7})
are replaced by their corresponding $k=0$ values, Eq.\ (\ref{eq:7})
transforms to Eq.\ (\ref{eq:1}). However, the motivation for replacing
$A(k)$ with $A(0)$ is not clear since this function [Eq.\
(\ref{eq:8})] decays on approximately the same length-scale as
$j_1^2(kR_1)$ in Eq.\ (\ref{eq:7}). It turns out that, if one employs
the identity
\begin{equation}
  \label{eq:9}
  \frac{S^L}{S^L+2A}-\frac{S^T}{S^T-A}=A\left(\frac{1}{S^T-A}+\frac{2}{S^L+2A}\right)
\end{equation}
and applies the ``continuum'' limit to the term in the parentheses, one gets
an alternative expression for the cavity field
\begin{equation}
  \label{eq:10}
  \frac{E_c}{E_0} = \frac{7(\epsilon+1)^2 + 8\epsilon }{12\epsilon(2\epsilon+1)} .
\end{equation}

The direct $k$-integration in Eq.\ (\ref{eq:7}) shows that the actual
solution branches between two continuum limits (Fig.\ \ref{fig:1})
through a singularity point which appears when a pole of the
longitudinal function 
\begin{equation}
  \label{eq:11}
  S^L(k^*)+2A(k^*)=0  
\end{equation}
falls on the real axis (Im$(k^*)=0$). Equation (\ref{eq:10})
accurately describes the cavity field at low polarities switching to a
solution close to Eq.\ (\ref{eq:1}) through a singularity. The
appearance of a real-$k$ singularity prevents us from using numerical
integration. The high-$\epsilon$ parts of the plots in Fig.\ \ref{fig:1} have
been calculated by residue calculus using the analytical
representation for $S^{L,T}(k)$ from the mean-spherical approximation
\cite{Wertheim:71} re-parametrized to give the exact $k=0$ limits in
terms of $\epsilon$ and $y$ \cite{DMcp:06}. The values of $\epsilon(y)$ have been
taken from MC simulations.

The real-axis singularity in Eq.\ (\ref{eq:7}) signals the appearance
of a non-decaying polarization wave induced by the cavity and radially
propagating from it through the entire liquid. This longitudinal
polarization wave is terminated at the boundary of a dielectric sample
where it creates surface charges. This picture is what we know as
macroscopic dielectric described by material Maxwell's equations for
which any field within polarized dielectric depends on the global
shape of the sample \cite{Boettcher:73}. This phase, which can be
described as conventional dielectric, is shown by the shaded area in
the space of parameters $\{\epsilon,R_1\}$ in Fig.\ \ref{fig:2}. For the
parameters in the un-shaded area (marked as ``polar liquid'' in Fig.\
\ref{fig:2}), the long-range polarization wave does not exist and any
polarization wave in the liquid decays on a microscopic length
\cite{SPH:81}. The local field is then independent of the sample shape
and the rules of macroscopic electrostatics do not apply. This regime
of relatively low polarities has an approximate solution given by
continuum limit of Eq.\ (\ref{eq:10}).

\begin{figure}
  \centering
  \includegraphics*[width=5.8cm]{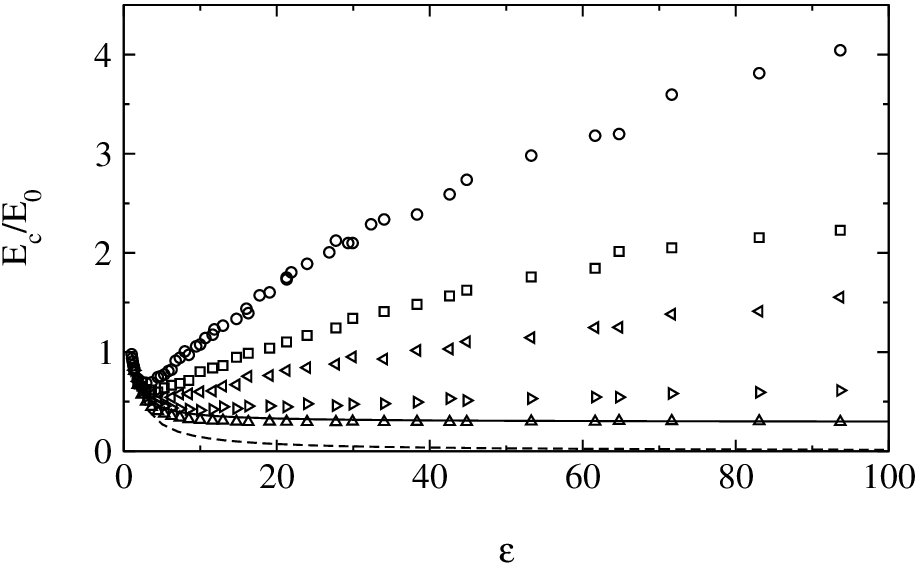}
  \caption{Cavity field obtained from MC simulations of the DHS fluid
    with a spherical cavity at the center of the simulation box. The
    points in the plot indicate cavities of varying size $r_1=R_0/ \sigma
    +0.5$: 1.0 (circles), 1.5 (squares), 2.0 (left-triangles), 3.0
    (right-triangles), 5.5 (up-triangles). Linear response
    approximation [Eqs.\ (\ref{eq:12}) and (\ref{eq:13})] was used to
    calculate $E_c/E_0$. The solid and dashed lines refer to continuum
    limits in Eqs.\ (\ref{eq:10}) and (\ref{eq:1}), respectively. }
  \label{fig:3}
\end{figure}

The quasi-continuum result of Eq.\ (\ref{eq:10}) in fact corresponds
to the separation of length-scales in which the cavity radius is
larger than the correlation lengths for both the longitudinal and
transverse dipolar correlations, $R_1 \gg \Lambda_{L},\Lambda_T$.  The use of Eq.\
(\ref{eq:9}) then largely eliminates the effect of the transverse
response on the continuum portion of the response function. The
transverse correlation length $\Lambda_T$ is an increasing function of
solvent dielectric constant \cite{Wertheim:71} growing to infinity at
the ferroelectric transition. Therefore, the semi-continuum limit
should become invalid at some $\epsilon$, and that happens through a
discontinuous branching of the continuum solution between Eqs.\
(\ref{eq:10}) and (\ref{eq:1}).  In order for a solution to switch to
the ordinary macroscopic limit, the singularity $k^*$ should be a part
of the sample's spectrum of wavenumbers. The spectrum of $\mathbf{k}$
is limited to a discrete set of lattice values for a finite-size
sample, and it is hardly possible for $k^*$ to coincide with one of
the lattice vectors. Indeed, when continuous $k$-integration in Eq.\
(\ref{eq:7}) is replaced with the lattice sum according to the rule
\begin{equation}
  \label{eq:11-1}
  \int d\mathbf{k}/ (2\pi)^3 \to L^{-3}\sum_{n,l,m} ,
\end{equation}
we do not observe a rising part of the cavity field (dash-dotted line
in Fig.\ \ref{fig:1}). In Eq.\ (\ref{eq:11-1}), $L$ is the size of the
cubic lattice and the lattice wavevectors are
$\mathbf{k}=(2\pi/L)\{n,l,m\}$. As expected from this calculation, we in
fact have not observed switching to the ordinary continuum in our
numerical simulations.

We have carried out MC simulations with the standard NVT Metropolis
algorithm, periodic boundary conditions, and the cutoff of the dipolar
forces at $L/2$ (see Ref.\ \onlinecite{DMjcp1:99} for the details of
the simulation protocol) . The initial configuration was set up as
face-centered cubic lattice with random dipolar orientations and
varied number of particles $N$. The spherical cavity was created at
the center of the simulation box and the solvent diameter was adjusted
to produce the bulk density $\rho^*=\rho\sigma^3=0.8$. Reaction-field corrections
with the dielectric constant equal to that of the liquid (from
separate MC simulations) were used for the dipolar interactions to
speed up the simulations. The results were identical to simulations
employing Ewald sums.

The cavity field was calculated from the linear response approximation
according to the equation:
\begin{equation}
  \label{eq:12}
  E_c/E_0 = 1 + (\beta/3)\langle\delta \mathbf{E}_s \cdot \delta \mathbf{M} \rangle - E_{\text{corr}} . 
\end{equation}
Here $\delta \mathbf{E}_s$ and $\delta \mathbf{M}$ are the fluctuations of the
field at the cavity center and the total dipole moment of the liquid,
respectively. The term $E_{\text{corr}}$, derived here from the procedure
suggested in Ref.\ \onlinecite{Neumann:86}, corrects for the cutoff of
the dipolar interactions at the distance $r_c$ in the simulation
protocol:
\begin{equation}
  \label{eq:13}
  E_{\text{corr}} = \frac{2(\epsilon-1)}{3\epsilon} \left(1+ \frac{\epsilon-1}{2\epsilon+1}\left(\frac{R_1}{r_c}\right)^3 \right) .
\end{equation}

Figure \ref{fig:3} shows the cavity field from MC simulations.  The
predictions of two continuum solutions, Eqs.\ (\ref{eq:1}) and
(\ref{eq:10}), are shown by the dashed and solid lines, respectively.
It turns out that both qualitative predictions of continuum
electrostatics [Eq.\ (\ref{eq:1})] are violated.  First, there is a
significant dependence of $E_c$ on the cavity size, as expected for
cavities comparable in size to the liquid particles.  Second, the
simulated dependence $E_c(\epsilon)$ never reaches the continuum limit of
Eq.\ (\ref{eq:1}), but instead levels off with increasing cavity size
at the solution given by Eq.\ (\ref{eq:10}).

\begin{figure}
  \centering
  \includegraphics*[width=5.8cm]{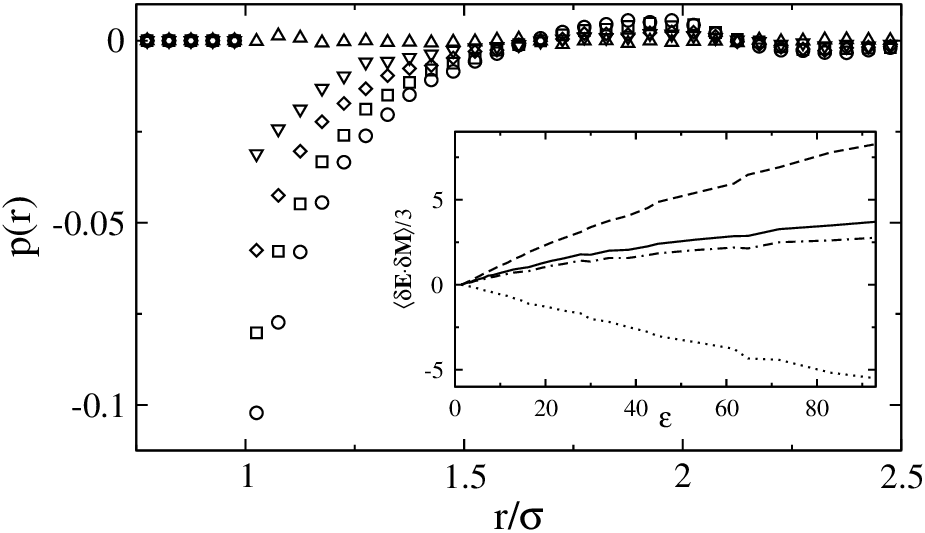} 
  \caption{Orientational order parameter $p(r)$ [Eq.\ (\ref{eq:14})]
    of the dipoles surrounding the cavity ($R_1/ \sigma =1.0$) for varying
    $\epsilon$: 1.4 (up-triangles), 8.5 (down-triangles), 17 (diamonds), 30.6
    (squares), 53.7 (circles).  The inset shows the splitting of the
    correlator $\langle\delta \mathbf{E}_s\cdot\delta \mathbf{M}\rangle$ in Eq.\ (\ref{eq:12})
    into contributions from the first solvation shell (dotted line),
    from the second solvation shell (dashed line), and first and
    second shell combined (dash-dotted line). The solid line indicates
    the overall correlator.  }
  \label{fig:4}
\end{figure}

The upward deviation of simulated cavity fields from Eq.\
(\ref{eq:10}) is a consequence of a particular orientational structure
on the cavity's surface.  Figure \ref{fig:4} shows the distance
dependence of the orientational order parameter formed by projecting
the unit dipole vector, $\mathbf{\hat e}_j$, on the unit radius
vector, $\mathbf{\hat r}_j = \mathbf{r}_j/r_j$:
\begin{equation}
  \label{eq:14}
  p(r) = \left\langle \sum_j P_2(\mathbf{\hat r}_j \cdot\mathbf{\hat e}_j)
            \delta\left(\mathbf{r}_j - \mathbf{r}\right) \right\rangle ,
\end{equation}
where $P_2(x)$ is the second Legendre polynomial. The surface dipoles
tend to orient orthogonally to the surface normal with increasing
polarity, a behavior well documented for 2D dipolar fluids
\cite{Weis:02}. Surface orientation of dipoles leads to overscreening
of the external field such that the electric field from the first
solvation shell is directed oppositely to the external field (inset in
Fig.\ \ref{fig:4}).  This effect is partially compensated by a
positive field from the second solvation shell, and it takes several 
shells to make the overall cavity field.  For larger cavities (not
shown here), the field of the first two solvation shells makes almost
the entire cavity field such that the solvent response is more local
and continuum-like.

In conclusion, we have followed the procedure, first suggested by
Maxwell \cite{Maxwell:v2}, to measure microscopic fields in a
polarized dielectric by carving cavities in it. The combination of
numerical simulations and analytical theory drew a new picture of what
is commonly called a macroscopic dielectric. We found that fields of
macroscopic electrostatics are formed only for sufficiently large
polarity of the liquid and the cavity size as a singularity in the
microscopic response function producing a non-decaying longitudinal
polarization wave. The total electrostatic free energy of polarizing
the dielectric \cite{Landau8} does not change at the branching point:
\begin{equation}
  \label{eq:15}
  \Delta F = -\frac{1}{2}\mathbf{E}_0\cdot 
         \left(\mathbf{M} + \mathbf{M}_c\right),
\end{equation}
where $\mathbf{M}$ is the total dipole of the polarized liquid. The
integrated dipole moment of the cavity $\mathbf{M}_c = - 3\Omega_0
\mathbf{P}/(2\epsilon+1)$ does not depend on which solution for the cavity
field is realized.  Therefore, the decrease in the electrostatic
energy of the cavity, caused by a lower cavity field, is released to
the longitudinal wave. The appearance of this solution within the
theory depends on the order of continuum ($R_0/(\Lambda_L,\Lambda_T)\gg
1$) and thermodynamic ($L\to \infty$ in Eq.\ (\ref{eq:11-1})) limits. It is
up to experimental measurements of cavity fields to determine which
limit should be taken first.

The cavity size reached in our simulations, $2R_0\simeq 5$ nm, is of the
order of that commonly realized for small nanoparticles, given the
typical length-scale of molecular liquids $\sigma\simeq 4-5$ \AA. We could never
reach the limit of macroscopic electrostatics on that length-scale
suggesting that electrostatics of nanocavities is not consistent with
material Maxwell's equations. Macroscopic electrostatics \cite{Landau8}
assumes that polarization is a continuous field terminated at the
interface where it creates a surface charge. The density of the
surface charge is equal to the polarization projection normal to the
surface \cite{Boettcher:73}.  Restructuring of the liquid interface,
in particular eliminating the normal polarization projection (Fig.\
\ref{fig:4}), will eventually affect the fields within dielectrics,
cavity fields included.  We need to notice that no direct measurements
of fields within microscopic cavities in polar liquids have been, to
our knowledge, reported in the literature.  Experimental evidence may
arrive from measurements of dielectric relaxation of photoexcited
dipolar impurities. 

This research was supported by the NSF (CHE-0616646).

\bibliographystyle{apsrev}
\bibliography{/home/dmitry/p/bib/chem_abbr,/home/dmitry/p/bib/liquids,/home/dmitry/p/bib/nano,/home/dmitry/p/bib/dm,/home/dmitry/p/bib/dielectric,/home/dmitry/p/bib/ferro,/home/dmitry/p/bib/solvation}

\end{document}